\documentclass[sigconf]{acmart}

\AtBeginDocument{%
  }

\setcopyright{acmlicensed}
\copyrightyear{2024}
\acmYear{2024}
\acmDOI{XXXXXXX.XXXXXXX}

\acmConference[IRESP 2024]{}{March 08--10 ,2024}{Shanghai, China}

\acmJournal{POMACS}
\acmVolume{37}
\acmNumber{4}
\acmArticle{CE202}
\acmMonth{8}

\acmSubmissionID{CE202}

\begin{document}

\title{VCR: Video representation for Contextual Retrieval}


\author{Oron Nir}
\orcid{0000-0002-8443-5014}
\affiliation{%
  \institution{Reichman University and Microsoft Corporation}
  \city{Herzeliya}
  \country{Israel}
}

\author{Idan Vidra}
\affiliation{%
  \institution{Reichman University}
  \city{Herzeliya}
  \country{Israel}}

\author{Avi Neeman}
\affiliation{%
  \institution{Microsoft Corporation}
  \city{Herzeliya}
  \country{Israel}
}

\author{Barak Kinarti}
\affiliation{%
 \institution{Microsoft Corporation}
 \city{Herzeliya}
 \country{Israel}}

\author{Ariel Shamir}
\affiliation{%
  \institution{Reichman University}
  \city{Herzeliya}
  \country{Israel}}
\orcid{0000-0001-7082-7845}
\renewcommand{\shortauthors}{Nir et al.}

\begin{abstract}

Streamlining content discovery within media archives requires integrating advanced data representations and effective visualization techniques for clear communication of video topics to users.
The proposed system addresses the challenge of efficiently navigating large video collections by exploiting a fusion of visual, audio, and textual features to accurately index and categorize video content through a text-based method.
Additionally, semantic embeddings are employed to provide contextually relevant information and recommendations to users, resulting in an intuitive and engaging exploratory experience over our topics ontology map using OpenAI GPT-4 (\href{https://github.com/oronnir/VCR}{GitHub}).
\end{abstract}




\begin{teaserfigure}
    \centering
    \includegraphics[width=0.85\textwidth]{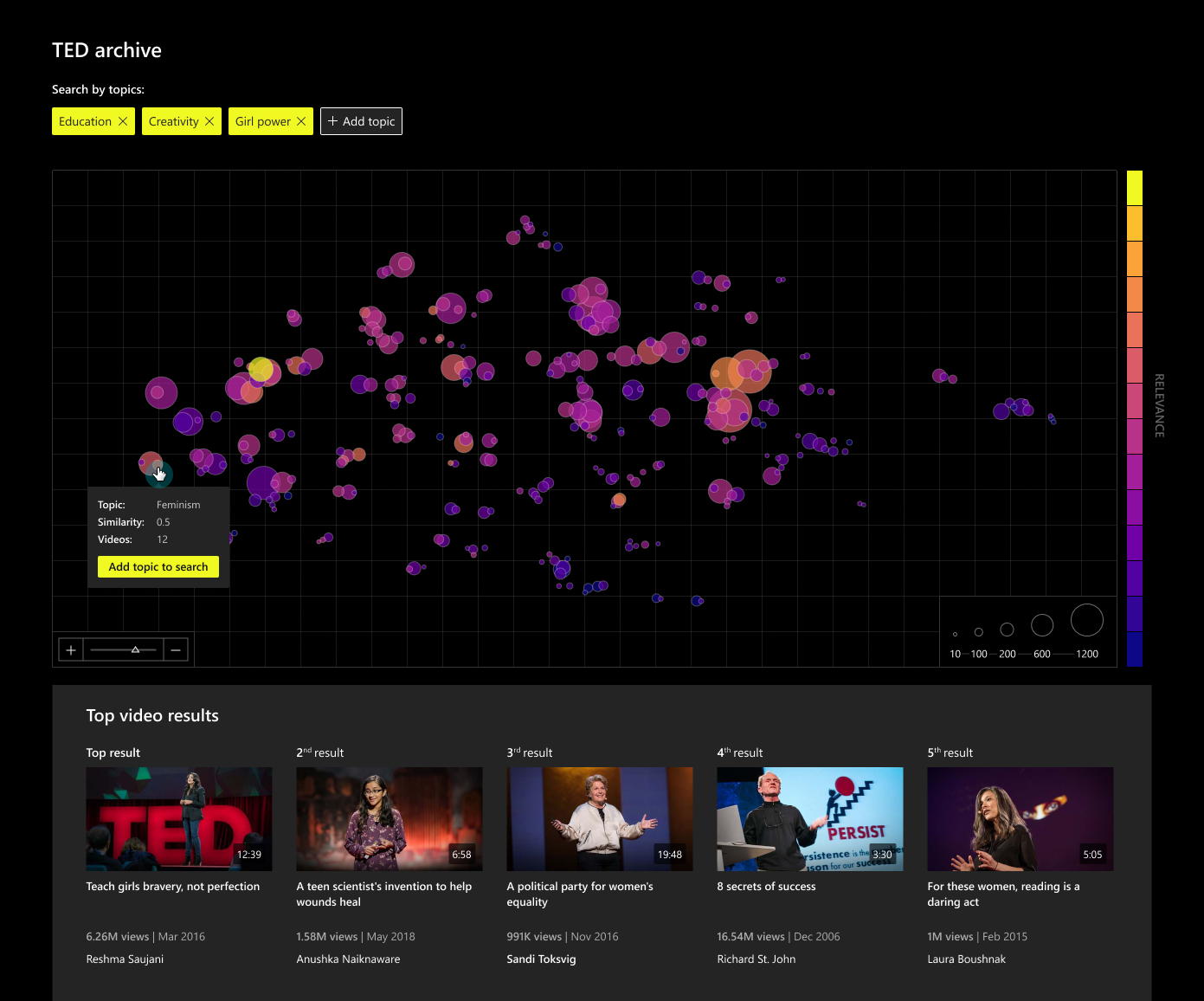}
    \caption{Video Archive Explorer empowers media users in content discovery scenarios while searching for a video with semantic similarity over multimodal insights. By browsing our Topics-Map users can tune their search while looking for semantically similar videos in the archive using OpenAI GPT4 and video insights embeddings. Insights extracted from TED.com.}
    \label{fig:explorer}
  \end{teaserfigure}

\maketitle

\section{Introduction}
Managing vast amounts of unstructured media content at scale poses a significant challenge for media and internet companies. Although content categorization by topics is an intuitive approach that facilitates content search, it is often deductive and may not appear explicitly in the video. For example, a video about ``Politics'' may not mention the word ``politics'', making content categorization an even more complex problem to solve. Manual tagging is a common approach, but it is costly, time-consuming, error-prone, requires periodic curation, and does not scale well. 
The success of data visualization user experiences lies not only in the design of the depiction but also often in the underlying data representation.
We propose an approach that employs multimodal video \emph{representation} and a \emph{visualization} tool based on topic detection and tracking in video (TDT) for a holistic cross-video semantic similarity exploration.

Videos convey information through multiple perception channels, including speech, visual text, actions with objects, scenery context, and individuals' identities, which all influence the message conveyed in the video. Hence, to properly address video indexing a multimodal approach is required. Multimodal machine learning algorithms have been widely researched to capture such patterns \cite{ngiam2011multimodal, chivadshetti2015content, qian2015multi}. Multimodality is also the approach that humans naturally leverage to capture the concepts in a video. However, videos are spread across various domains, including web video, social media, entertainment, and more, and each type delivers information in a different mixture of channels. For instance, educational content and news broadcasts typically convey heavier visual text information than other genres, while featured films and entertainment typically feature famous people and conversational data. 

In this paper, we focus on high-information density videos and generate a semantic representation that captures their main themes. We targeted organizational, educational, and entertainment media content types due to their high information density levels in automatic speech recognition (ASR) transcription, Optical Character Recognition (OCR), and contextual visual captioning. 
We evaluate our proposed methodology on the \textsc{TED} repository by the TED organization \cite{ted2023} and using the \textsc{MSR-VTT} dataset~\cite{xu2016msr}. 
Still, the proposed methodology could be easily extended with additional sources to enable the recognition of new video types for different domains, such as animal recognition for nature films.
The \textsc{TED} dataset, we suggest, was obtained from publicly available TED talks online. The model takes as an input the video transcription, optical character recognition, and frame captioning from both textual and visual information, and is evaluated on various tasks.

We propose an interactive media archive explorer user experience we call \textit{Topics-Map}. It is designed to guide the user through the concepts conveyed in its media archive. We utilize topics ontology visualization and semantic representation in 2D. The main search functionality is based on the set of selected topics. Thus, the map enables an intuitive adjustment to the query while refining the search to discover related videos.

Our content discovery approach divides into offline indexing and the online querying mechanism - the Topics-Map.
The offline step runs the aforementioned models on the archive.
To fuse different video insights together we decode all details into a time-based interleaved textual format and encode it again using Large Language Models (LLMs) into a latent semantic representation.
At inference time, the user specifies a set of topics, using the Topics-Map, while the system leverages GPT-4 LLM to generate a comprehensive video description that depicts the type of video the user would like to explore. This textual description is embedded using LLM and efficiently ranks the entire archive according to the query. This process yields a ranked list of relevant videos (Figure~\ref{fig:explorer}).

The key contributions of this paper are: (1) The development of a novel methodology for multimodal text-based video representation using a LLM with SoTA results on unsupervised Text-Video Retrieval task over \textsc{MSR-VTT}, (2) A visualization tool that allows effective content discovery on a large media archive - using OpenAI GPT-4 and state-of-the-art text encoders, (3) the \textsc{TED} multimodal text-based dataset and \textsc{MSR-VTT}~\cite{xu2016msr} multimodal text-based dataset. We provide a full evaluation of the proposed methodology on the \textsc{TED} dataset, a user study, and the code.
Overall, this paper provides a new visual approach for exploring video archives that can help organizations transform their media content into business value more efficiently and effectively.

\section{Related work}
Topic modeling (TM) is a well-established field in natural language processing (NLP) that has been widely explored for TDT. TM dates back to the development of pLSA/I ~\cite{hofmann1999probabilistic} and LDA ~\cite{blei2003latent}, which both took a generative probabilistic approach for word-topic vs. topic-document distributions. Other significant developments in this field include Hierarchical Dirichlet Processes~\cite{10.1198/016214506000000302}, Word2Vec~\cite{NIPS2013_9aa42b31}, and GloVe~\cite{pennington2014glove}, which paved the way for subsequent neural network architectures. 
Multimodal TDT have been explored  with various methods for feature fusion including classification-based fusion, clustering-based fusion, and online TDT feature fusion. E.g., Virtanen et al.~\cite{virtanen2012factorized} proposed a hybrid approach that integrates LDA and canonical correlation analysis to embed different modalities and classify topics simultaneously. Harakawa et al.~\cite{harakawa2018tracking} applied a clustering-based approach using graph clustering with Modularity Maximization.~\cite{li2016online} developed an online-based TDT approach using a learning algorithm that incrementally adapts to reveal new topics.

Recurrent Neural Networks (RNNs) have been used for document summarizing and classification e.g. SummaRuNNer~\cite{nallapati2017summarunner} which uses GRU~\cite{chung2014empirical} to summarize texts. As RNNs have been challenged to avoid the vanishing and exploding gradients problem over the years, even with advanced works like ELMo~\cite{peters2018contextualized}, other architectures based on the Attention mechanism have managed to significantly push the boundaries of LMs with the Transformer~\cite{vaswani2017attention} architecture.
The BERT~\cite{devlin2018bert} architecture leveraged the Transformer mechanism to encode semantics. RoBERTa~\cite{liu2019roberta} has optimized the same architecture for several tasks. BART~\cite{lewis2019bart} is one of many strong incremental works that followed it. 
Another important challenge towards our goal is long texts classification. The LM's capacity poses a challenge while working with long texts like video ASR. This difficulty is addressed in many ways e.g. Longformer~\cite{beltagy2020longformer}, Big-Bird~\cite{zaheer2020big}, Hierarchical DocBERT~\cite{adhikari2019docbert} and others. Recent developments by OpenAI have introduced the generative pre-trained transformer (GPT) family of LLMs~\cite{radford2019language, openai2023gpt4} which introduced ground breaking advancement in language understanding and generation.

Another challenging aspect is the size of the label set. Chang et al.~\cite{chang2020taming} have analyzed the extreme multilabel text classification (XMC) problem on product descriptions queries online for a retail website. They have demonstrated their work, X-Transformer, on MNLI GLUE~\cite{wang2018glue} with $2K$ labels, on Wiki-500K~\cite{varma2019extreme} with $500K$ labels, as well as on XMC with up to $1M$. The reported metrics were precision@$k$ and recall@$k$: $k \in \{1, 3, 5, 10\}$. X-Transformer's architecture was designed to avoid limiting the representation capacity by using prior adaptation methods like ELMo~\cite{matthew2018peters} and not LMs thus excluded from this analysis.
One powerful LM architecture that followed the aforementioned works is Decoding-enhanced BERT with Disentangled Attention or DeBERTa~\cite{he2020deberta}. Its followup paper V3~\cite{he2021debertav3} has significantly improved its performance using ELECTRA-Style Pre-Training with Gradient-Disentangled Embedding Sharing.

Large video archive retrieval is a basic functionality at the core of any media archive explorer. TRECVID~\cite{chaisorn2010trecvid} have suggested a method and the known-item search (KIS) task for locating a single true answer for a textual query over a video archive in a multimodal search. SEA was suggested for a text-based search over multimodal set of features leveraging a sentence encoder for Ad-hoc search \cite{li2020sea}. The evaluation task was Ad-hoc Video Search (AVS). The main difference between the two tasks lies in the KIS assumption for a single correct answer while AVS accepts more.
Text-Video Retrieval is the formal KIS task performed today over datasets like \cite{xu2016msr, hendricks18emnlp, rohrbach2019large, miech19howto100m, huang2020movienet, bain2020condensed, soldan2022mad} with VAST~\cite{chen2023vast} being the SoTA on many of them. SoTA methods today use all relevant modalities e.g., audio, video, and captions, while depending on an over-fitting phase towards a dataset and a downstream task.

Large scale media archive corpus visualization is an important area of research that has been gaining more attention in recent years. With the ever-growing amount of multimedia data available, it has become crucial to develop effective methods for exploration and analysis. One of the recent efforts in this direction was conducted by Wu et al.~\cite{wu2018multimodal}. John et al.~\cite{john2019visual}, who proposed a method based on topic inference and clustering for visualizing multimedia news data. Renoust et al.~\cite{renoust2021multimedia} suggested a featured films archive browser using a multilayer network modeling approach that was validated over the IMDb dataset. These studies demonstrate the importance of large scale media archive corpus visualization for exploration and analysis of multimedia and provide valuable insights into the potential of different approaches towards this goal.

Several other studies have developed large-scale media archive exploration visualization tools. Television broadcast archive visualization applications \cite{michael2009visualizing, haesen2013finding, muhling2022viva}. Others have developed visual analysis systems for exploring large-scale multimedia news archives using a combination of topic modeling and visualization techniques as well as a facial recognition based politics TDT \cite{renoust2016faceJap, ren2019evaluating}.
Nir et al.~\cite{nir2022cast} have suggested applications for animation. Shahaf et al.~\cite{shahaf2013metro} have structured storylines visualization of metro-map narratives.
A recent paper by Afzal et al.~\cite{afzalvisualization} conducts a wide survey on media corpus visualization. One interesting work is VIAN ~\cite{halter2019vian} which suggested a novel artistic color control through a user experience which helps post production analysis and editing of color according to its semantics. VisTa~\cite{fan2019vista} was suggested for assisting in video tagging. Other related domains included sports broadcast, image collages, surveillance footage visualization, and more.

In recent years, online education through Massive Open Online Courses platforms, like \textit{Coursera}, have gained immense popularity \cite{shi2015vismooc}. Platforms that offer these courses are interested in analyzing engagement with the course material through web access logs or clickstream data \cite{chen2018viseq}. Visual analytics tools, such as PeakVizor~\cite{chen2015peakvizor}, enabled experts to gain insights from raw data. 
He et al.~\cite{he2018vusphere} developed a visual analytics system, VUSphere, for the exploration and comparative analysis of video utilization in courses, students, and distance-learning centers. The system provided multiple interactive views of the video utilization, such as video access and completion rate, to support the identification of useful insights.
We aim to empower users towards semantic content discovery using a text-based video insights fusion and LLMs.

\section{Multimodality}

\textbf{Automatic speech recognition (ASR)} takes a raw audio file and recognizes the spoken text. This classic task have surpassed human parity with the work of \cite{xiong2016achieving} and therefore this is the service used to extract spoken text with in this work. We believe that the conversations in the video are a strong video signal for different media types e.g., in Education.

\textbf{Optical character recognition (OCR)} is a classic computer vision task that extracts the visual text appeared in an image. OCR tracking in videos requires higher level of recognition to match permutations of the same text over multiple frames where it may vary due to recognition errors, occlusions, and other effects. To overcome these challenges and yield a corrected result per video, a consolidation algorithm was applied using multi-sequence alignment~\cite{wemhoener2013creating} to cluster and correct unique texts out of all permutations in an unsupervised manner. The underlying hypothesis of the OCR signal usage to represent video topics could be easily explained with examples like a presentation’s deck, the titles on a news broadcast, the rolling credits of films, or even social media content. All examples above describe a natural information communication that we humans easily perceive hence, OCR was selected as a source signal. The underlying method is based on Florence~\cite{yuan2021florence}.

\textbf{Frame Captioning} has been an active research area in computer vision, and recent advancements have shown promising results in generating accurate and meaningful captions for images. Florence~\cite{yuan2021florence}, a SoTA image captioning model, utilizes a transformer-based architecture to generate captions that capture fine-grained details and contextual information. However, the application of image captioning models to video frame description poses unique challenges, as it requires the generation of temporally coherent and contextually relevant captions that convey the narrative of the video. One of the key challenges is to maintain the continuity of the context across multiple frames, as the temporal relationships among the frames can be complex and non-linear. Another challenge is to balance the trade-off between generating detailed and accurate descriptions for individual frames vs. producing a coherent and concise narrative for the entire video. To address these challenges, we aim at leveraging a LLM to fuse the multimodal information.

\section{Method}

\begin{figure*}[tbp]
  \centering
  \includegraphics[width=\linewidth]{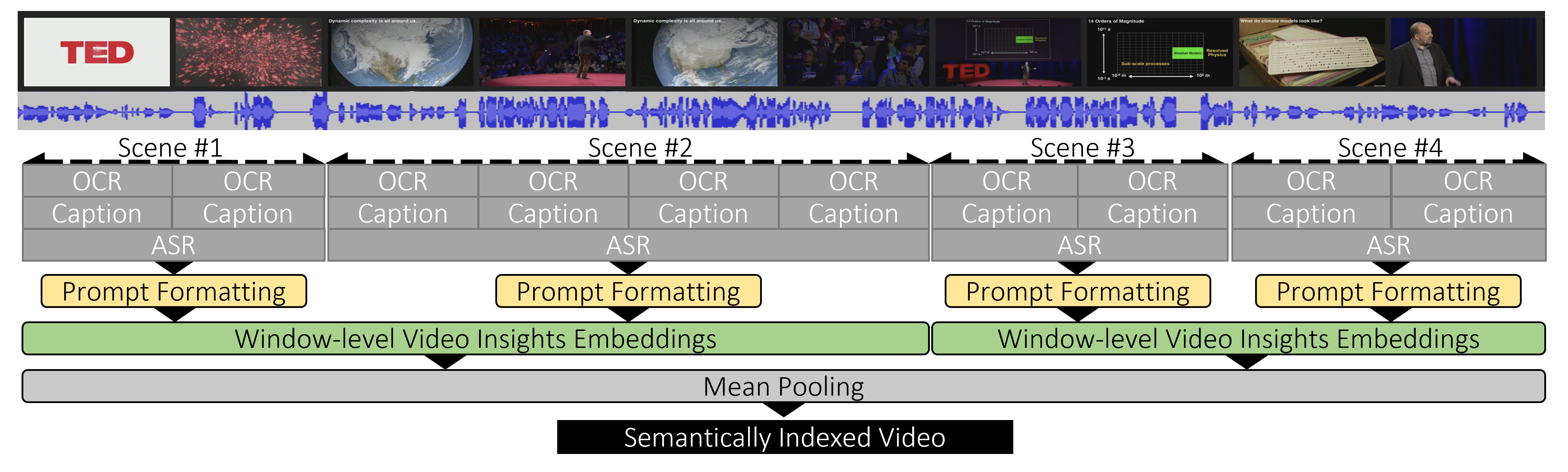}
  \caption{Our Text-based multimodal video representation with LLMs.}
  \label{fig:MethodArchitecture}
\end{figure*}

In this section, we first present our approach for content-based video retrieval using Ad-hoc Video Search (AVS)~\cite{li2020sea} on multiple AI models and then describe the interface. 
Our method is designed to understand conveyed information in media files per channel by leveraging machine vision and audio analysis algorithms. We then aggregate these video insights into a semantic vector representation (see Figure~\ref{fig:MethodArchitecture}). Specifically, we utilize different LLMs to embed the interleaved content prompts in a latent semantic space which serves as the video index. To enable efficient content discovery, we propose an application that allows users to search the video archive by embedding an auto generated language query from selected topics on the Topics-Map, which is then used to rank the search results with Cosine Similarity. We compare both a supervised learning approach on language models such as BERT, RoBERTa, and DeBERTa, as well as leveraging the popular GPT-based OpenAI embeddings.

\subsection{A supervised approach}
To represent the recognized video insights semantically we first fine-tune a LLM to classify these insights according to the video topics using Multilabel sequence classification and Binary Cross Entropy. The hypothesis behind selecting the topics of the video as the pretext task is since topics are by definition the holistic subject of the video which makes a suitable trait while trying to distinguish videos or look for content which is semantically similar.

During training, a sampling method was used to fit the backbone LLM towards representation. First, an initialization step, data indexing mapping labels to documents. Next, a batch sampling method, sampling a label, a document, and a $512$ token window. Using this data loader method, a text segment is sampled in constant time and the window size is kept constraint to avoid long text memory inefficiency. Its hypothesis is that any sampled window holds the same global multilabel topics.

Our approach takes a textual representation of insights ordered by the time of appearance and name the main topics of the video. We compare different backbone architectures: BERT~\cite{devlin2018pretraining}, RoBERTa~\cite{liu2019roberta}, and DeBERTa~\cite{he2021debertav3}. Each LLM is trained on a random text segment of 512 tokens and evaluated on the mean pooling of sliding windows of the same size without intersection.

\subsection{A learning-free approach}
The recent development of GPT has shown significant advancements in NLP. GPT is a LLM that has been pre-trained on a massive corpora of diverse texts, making it capable of generating high-quality text outputs with coherent and fluent language. Besides its capabilities in text generation, GPT can also be used as an encoder to extract features from textual input. GPT-based encoders have shown promising results in various NLP tasks such as text classification and question-answering. GPT can also be used as an encoder-decoder architecture for text generation. This approach have shown success in various applications like chatbots, translation, and summarization \cite{radford2019language}.

Like the supervised approach, we first run the aforementioned video AI models on the entire archive. Then we split the extracted insights into semantically coherent segments of up-to $4,096$ tokens using a scene segmentation algorithm. Then, we serialize all insights to text, order it according to its start time and tag it with the source type, e.g., ``[OCR] \textit{<some visual text on frame>}''. These interleaved segments are then encoded using the OpenAI encoder and indexed in an index matrix of size $N \times M$ where $N$ is the number of archive segments and $M=1,536$ is the embedding dimension.

\subsection{Topics ontology query generation}
The browsing experience described in details in the following section from the user experience perspective. However, its level of abstraction is enabled by auto generation of detailed textual queries using GPT-4~\cite{openai2023gpt4} with prompt engineered request including a set of topics provided by the user.
At inference time, when the user is changing the set of selected topics, a new query string is generated with the prompt below (see section~\ref{sec:implementation_details}). It is then rephrased by GPT-4 as a description paragraph which is encoded and applied with a Cosine Similarity on the index matrix while returning the \textsc{K} nearest neighbors ranked by relevance.

\subsection{The Topics-Map}
The Topics-Map represents the different ontology topics on a 2D plane using semantic representation of their names which resides concepts like ``\textit{Law}'' closer to ``\textit{Policy}'' than topics like ``\textit{Biology}''.
Its motivation comes from the hypothesis that users prefer semantic order rather than lexicographical order when exploring an ontology due to its natural association.
The Topics-Map positions are kept fixed to minimize the user's orientation overload.
Color coding the ontology w.r.t. the set of selected topics is done using our LLM for embedding both the auto-generated query against the entire ontology and ranking using Cosine-Similarity.

\subsection{Implementation Details}
\label{sec:implementation_details}
Training the LLMs was done by first training two epochs while freezing the encoder and training only the classifier with a learning rate of $10^{-3}$, later additional 38 epochs on the entire network using Nvidia A100 GPUs with a batch of size 32 and learning rate: $10^{-5}$. 
We used \textit{AdamW} optimizer. The application of the model is done by taking the last linear layer before the classifier. When the input is longer than 512 tokens, the model runs a sliding window with no intersection and uses mean pooling for representation.

OpenAI embeddings was used with the ``text-embedding-ada-002'' engine. Long input texts were invoked and aggregated with Mean Pooling.
We use the following prompt to inflate the set of selected topics as the seed for a textual query using OpenAI's GPT-4: ``\textit{Write a full description for this TED talk which discusses the following topics <selected topics>.}''. While the user is required to pick a small set of concepts, the generative model fuses these ideas together to form a detailed textual query.

\section{User Experience Design}

For content discovery purposes, we propose an interactive UX which is built out of three main components: (1) a set of selected topics from either a closed set or custom terms, (2) the Topics-Map, and (3) the top results panel. 
The flow begins with topics selection, either from the ontology or custom topics, which is automatically augmented into a full video description and embedded to color code semantic similarity as described in section~\ref{sec:implementation_details}. By hovering the Topics-Map a tool-tip reveals the semantically mapped topics with attributes like the topic's relevance to the current query, the frequency of the topic in the dataset (also coded as the size of the circle) and its name. Any change to the set of selected topics yields a new top video results panel as well as an updated color code of the Topics-Map. The location and the size of the circles remain frozen.
The semantic space is created either by a \textit{supervised} or a \textit{learning-free approach}. It preserves local ontology similarities using dimensionality reduction on the 2D plane. Each circle on the scatter plot represents a topic, and users can either interact with the plot to navigate the Topics-Map or to refine its topic set selection. The underlying hypothesis is that interacting with locally related topic suggestions on a 2D grid can assist the user in correcting a query to its true intent.
This interface provides an intuitive and effective method for content discovery, leveraging the power of semantic embeddings for quick and efficient exploration of large-scale video corpora. 
There are four underlying steps in the application implementation. Initially the user is (1) adding topics which are used to generate a prompt to GPT-4. (2) GPT-4 generates the video description which is then (3) embedded and correlated with the entire ontology and (4) updates the Topics-Map. This process is repeated indefinitely.

The proposed interactive UX has several advantages over traditional browsing methods. First, the combination of recommendations and a color-coded 2D Topics-Map provide a holistic approach to video discovery by allowing users to explore content that they didn't consider before. The map serves as a visual representation of the video archive, allowing users to easily navigate through different topics. Additionally, the incorporation of both exploration and exploitation search strategies provide users with a more personalized browsing experience. Users can switch between the two strategies based on their current needs and preferences, leading to a more efficient and satisfying video browsing experience. 
By focusing the user's attention on the Topics-Map we steer away from technical challenges like big data visualization as the user interface requires to visualize the ontology and not the entire archive. Projecting the ontology on the map is done with t-SNE \cite{vanDerMaaten2008} reduction of topic names' embeddings.
Overall, this approach offers an engaging and immersive user experience that encourages discovery and personalization while emitting cognitive load through the Topics-Map abstraction.

The Top Video Results panel displays relevant metadata on the ranked video results. The visualized information include the video title, thumbnail, author's name, the event date, and the number of views and likes online. Clicking the thumbnail opens a video player with the relevant TED talk. The search interface allows users to update their search and provide near real-time feedback to refine the query based on the video archive's content.

Our user experience is designed for ongoing session with a user to continuously refine the query and switch between exploration and exploitation strategies trading off precision and recall during the search session. Initially the map is grayed out until the first topic is selected.
As the user types the query significant topics are highlighted and the top results on Figure~\ref{fig:explorer} are updated.

\section{Evaluation}

The method was evaluated in three steps. We first compared different architectural backbones, then ASR-only vs. multimodal video insights, and conclude with quantifying the impact of the method as a recommendation/search engine like other known-item search (KIS) tasks.
Three datasets were leveraged in these experiments. The \textit{TED} dataset, \textit{TDT2 English Audio} Topics dataset~\cite{cieri1999tdt}, and \textit{MSR-VTT} dataset~\cite{xu2016msr}.

\textbf{TED Topics Dataset}\newline
We have introduced and open sourced the \textsc{TED} dataset for Multilabel Video Topics classification. The dataset is based on TED.com website which assigns a variable size of label-set per talk. These were considered as the target variable of this task. 
The dataset consists of $5,439$ \textsc{TED} talks which were originally represented by $345$ topics. The topics have been distributed in a long-tail distribution having ``Science'' as the most popular topic with $1,215$ talks while ``Hinduism'' for instance, appears in a single talk.
We've decided to keep topics which have at least $10$ talks and lose TED specific topics which are not generalize-able to other domains, e.g., ``TEDx''.
Splitting the dataset into \textsc{Training} (80\%), \textsc{Validation} (10\%), and \textsc{Test} (10\%) was done according to a method for multilabel balance preserving stratified sampling~\cite{sechidis2011stratification}.
The \textsc{TED} organization provides also the textual descriptions of character length $\mu=369, max=1,391$. These descriptions are short implicit characterization of the content. We used them later to mimic a textual query by users for which the suggested method would try and recommend from a large set of videos. We believe that its insinuated articulation makes a good query which a user would have used to find some specific video.

\textbf{Augmented TDT2 Topics Dataset}\newline
To finetune our model towards both multilabel classification and document representation we leveraged \textit{TDT2}~\cite{cieri1999tdt}.
TDT refers to automatic techniques for finding topically related material in streams of data such as newswire and broadcast news. 
We have augmented this document dataset with online articles from open forums and manually label them according to the \textit{IPTC Media Topics} ontology~\cite{iptcmediatopics}. The dataset contains over 200k documents and the topics ontology holds 4k categories. This ontology deals with different concepts from Politics, Commerce, Tech, Society, Healthcare, Education, and essentially top media-related titles. The dataset was split into training (80\%), validation (10\%) and test (10\%) with the same method~\cite{sechidis2011stratification} and used in our supervised approach.

\textbf{Topics Representation}\newline
To evaluate the models' performance on our test set we considered the video topics multilabel classification task as our pretext task with standard multilabel metrics like Precision, Recall, and F1 score in its ``Micro'' version of the metrics as we analyzed the entirety of TED and maintain label ratios between dataset splits.
We trained and compared the backbones BERT~\cite{devlin2018bert}, RoBERTa~\cite{liu2019roberta}, and DeBERTa-v3~\cite{he2021debertav3} through multilabel classification experiments on \textsc{TED} test set (see results on Table~\ref{tab:TedAblationClosed}). 
To compare apples-to-apples we used a ``base'' model version across backbones i.e., $~86M$ parameters.
The best F1 results were achieved using DeBERTa-v3 using finetuning upon our augmented \textit{TDT2} Topics Dataset.
We split the result w.r.t. two windowing strategies, the first takes only the first window while the later runs a sliding window with stride size equal window size equal 512 tokens. Our model has out-performed vanilla DeBERTa-v3-base on this task by a significant margin on all measures.

\begin{table}[ht]
\center
  \centering
    \begin{tabular}{p{0.28\linewidth} p{0.26\linewidth} | p{0.06\linewidth} p{0.1\linewidth} p{0.08\linewidth}}
    \toprule
    Backbone & Window strategy & F1 & Precision & Recall\\
    \midrule
    BERT          & First & 17.7 & 36.4 & 13.0\\
    RoBERTa       & First & 43.8  & 45.2 & 42.4 \\
    DeBERTa-v3    & First & 41.0  & 48.4 & 35.5 \\
    Ours (refined Dv3) & First & \textbf{60.4}  & \textbf{72.9} & \textbf{51.6}\\
    \hline
    DeBERTa-v3    & Sliding & 40.6 & 56.9  & 31.5 \\
    Ours (refined Dv3) & Sliding & \textbf{65.0} & \textbf{66.6}  & \textbf{63.6} \\
  \bottomrule
  
\end{tabular}

  \caption{Ablation study on \textsc{TED} Topics test set (N=542) using ASR input on 'base' class backbones and window strategies.}
  \label{tab:TedAblationClosed}

\end{table}

\begin{figure}[tbp]
  \centering
  \includegraphics[width=\linewidth]{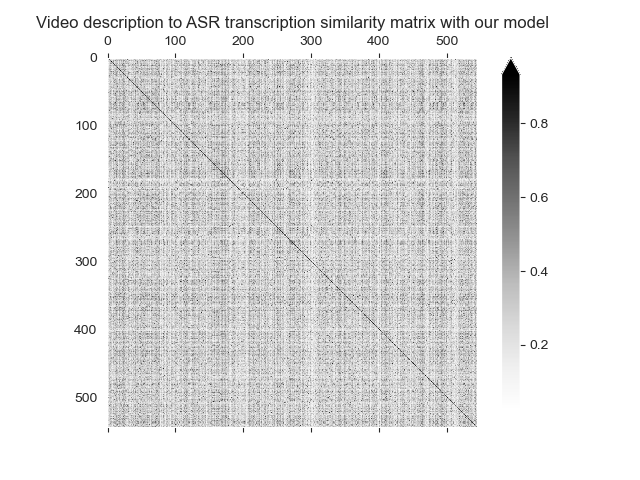}
  \caption{Cosine-Similarities between TED descriptions and its corresponding ASR transcripts. The bold main diagonal illustrates the significance of our method in matching them.}
  \label{fig:RanksCorrelationMat}
\end{figure}

\textbf{Search Recommendations}\newline
To further measure the application's recommendations the entire test set was embedded. A ranked list was recommended per video description by its Cosine-Similarity. The quality of recommendations were quantified using Mean Reciprocal Rank (MRR)~\cite{jurafskyspeech} and $Recall@k$.
The MRR is a measure to evaluate search engines that returns a ranked list of answers to queries. For a single query, the reciprocal rank is $\frac{1}{rank_i}$ where $rank_i$ is the position of the first correct answer. If no correct answer was returned in the query, then the reciprocal rank is 0. For multiple queries Q, the Mean Reciprocal Rank is the mean of the Q reciprocal ranks:
$MRR=\frac{1}{|Q|}\displaystyle\sum_{n=1} ^{|Q|} \frac{1}{rank_i}$. We used $Q=N=542$ hence, the score was bounded to the range [$\frac{1}{542}$, 1] where a value of $1$ indicates that the first recommendation is always correct. If the right result is always ranked second its MRR is $\frac{1}{2}$ even for a large $Q$ which is a strict penalty.
Given a \textsc{TED} video description, a correct answer was considered to be its corresponding video.
Although no model was trained on the descriptions, We report the MRR on the \textsc{TED} Test dataset.
The correlation matrix between all video descriptions and all video transcriptions is illustrated in Figure~\ref{fig:RanksCorrelationMat}. The main diagonal's significance highlights how our method successfully ranks the relevant videos first out of the \textsc{TED} Test set (N=542). The resulting $MRR=0.645$ which means that the mean rank of the correct video using our method is $1.55$ out of $542$ possibilities.

The same experiment was conducted using multimodal video insights mentioned above. Our model has reached $63.9$\%, $52.2$\%, and $57.4$\% Precision, Recall, and F1 Micro respectively on the Multilabel Classification task. While embedding the TED Test set along with its description the method has reached an MRR of $0.727$ i.e., exceeding ASR-only performance by $13$\%. The mean rank of the correct answer according to our method is $1.37$ out of $542$.
$Recall@k$: $R@1=54.8$\%, $R@3=70.7$\%, $R@5=77.5$\%, $R@10=85.4$\%.

\textbf{OpenAI Embeddings}\newline
To evaluate the OpenAI embeddings we leveraged the \textsc{TED} descriptions as queries per talk again. We quantified the effectiveness of our method by reporting MRR using the OpenAI embeddings, meaning without training. The OpenAI text representation provides a vector of dimension $1,536$ which encodes the latent information in the input text.
By using these embeddings as the video representation we've reached $MRR=0.950$ on the ASR-only input and $MRR=1.000$ on the multimodal video insights input respectively.
Perfect score that practically solves the task. See comparison in Table~\ref{tab:MRR}.
Base models of $86M$ parameters are not comparable to GPT3.5 scale for which the true number of parameters is kept unknown to this moment but estimated in $175B$ which is $2k$ times more. In addition, its rich expressiveness ($1,536$) is significantly larger than ours ($314$).
The key takeaway from these experiments is the contribution of multimodality to the video representation.
When evaluating the MRR with OpenAI's embeddings on the entire TED catalog (N=$5,439$) we reach $MRR=0.871$ on ASR-only and $MRR=0.999$ on the multimodal version.
To this day OpenAI did not share its training data hence we cannot guarantee \textsc{TED} was not included in some way. In-fact, a simple interaction with ChatGPT indicates that the system knows something about \textsc{TED} thus, confirms some level of training set contamination. Yet, the text-based multimodal superiority remains a significant finding for this task.

\begin{table}[ht]
\center
  \centering
    \begin{tabular}{p{0.32\linewidth} | p{0.32\linewidth} | p{0.23\linewidth}}
    \toprule
    Embeddings/MRR[$\uparrow$] & Ours (refined Dv3) & OpenAI \\
    \midrule
     Transcription         & 0.645 & 0.950 \\
    \hline
    Multimodal        & \textbf{0.727} & \textbf{1.000} \\
  \bottomrule
\end{tabular}
  \caption{An ablation study over the \textsc{TED Topics} test dataset (N=542) comparing our model with OpenAI GPT Embeddings.}
\label{tab:MRR}
\vspace{-5pt}
\center
  \centering
    \begin{tabular}{p{0.2\linewidth} p{0.14\linewidth} | p{0.07\linewidth} p{0.07\linewidth} p{0.07\linewidth} p{0.07\linewidth} p{0.07\linewidth}}
    \toprule
    Model         & Signal & MRR & R@1 & R@3 & R@5 & R@10\\
    \midrule
    Dv3 (No FT)  & ASR    & 0.189 & 12.7 & 18.9 & 23.8 & 31.6\\
    Dv3 (No FT)  & MM     & 0.229 & 14.6 & 24.6 & 31.5 & 40.1\\
    \midrule  
    OpenAI        & ASR    & 0.467  & 41.0 & 48.0 & 51.5 & 56.2 \\
    OpenAI        & MM     & \textbf{0.634}  & 54.3 & \textbf{68.6} & \textbf{73.9} & \textbf{81.1} \\
  \bottomrule
    SoTA \cite{chen2023vast} & MM & NA     & \textbf{63.9} & NA   & 68.3 & 73.9\\
  \bottomrule

\end{tabular}
  \caption{MSR-VTT MRR and Recall@k with no fine-tuning.}
  \label{tab:MsrVtt}
\end{table}
  
\textbf{MSR-VTT: A learning-free evaluation}\newline
Our method was evaluated on the \textsc{MSR-VTT} \cite{xu2016msr} Test dataset (N=3k) Text-Video retrieval task without finetuning. The domain gap is significant as the average video is 80 times shorter than TED. The resulting metrics on Table~\ref{tab:MsrVtt} underlines the significant advantage of our multimodal approach which is on par with SoTA with no training.

\section{User Study}
As the Evaluation was focused at a KIS task of a single true video retrieval, we focused the user study on AVS tasks to demonstrate the UX recall orientation.
The goal was to quantify the system's effectiveness in enabling content discovery and exploration.
We considered $k=5$ to be a reasonable item set to retrieve at a time. To cancel learning curve we provided the subjects with a warm-up demonstration. To cancel order bias we alternated the order of methods between our method and the baselines.
The user study included ten participants at ages 25--60 from various backgrounds which are familiar with graphical user interface basic usage on a personal computer. No reward was offered to volunteers.
The interview was designed to first align on the suggested method with the user's expectations, conduct two content discovery tasks and collect both performance metrics as well as qualitative evaluation on the user experience. An overall interview duration took between 60 to 70 minutes.
Each participant have taken six search sessions in total corresponding to testing performance on our approach vs. the baselines in two types of tasks.

The participants were asked to describe their experience with media archive search and content discovery. Most of them described tutorial video search on services like Google Search, Bing, YouTube, and Social Media platforms. Participant\#6 have mentioned internal corporate media repository tool named Amazon|knet, which she uses for technical training. Some mentioned home use of Netflix and other VoD streaming services, PC file explorers on Mac and Windows as well as storage services like Dropbox, GDrive, and OneDrive.
One participant (P\#2) have mentioned also CCTV footage browsing in his day job. With 20 years of experience in security media analytics he is considered as Expert1.
Participant\#8 manages the video library of a large university with all Zoom course lectures using Kaltura. With over 30 years of experience as the prior CTO of the National Library and COO in different communication corporates, he's considered as Expert2.

The \textsc{TED} website supports three ways to search. \textit{TED.com/topics} is an alphabetic ontology index in a lexicographic order. It supports a single topic selection. \textit{TED.com/talks} is a structured query interface which allows users to filter talks by topic names up to two topics at a time. The third web page is the \textit{TED.com/search} component which runs a query over textual indexed set of insights like title, speaker name, topics, transcript and more. Since the later is the strongest option for content discovery we considered it to be the \textsc{TED} baseline.
\textit{youtube.com/@TED/videos} is the TED YouTube channel which provides a search box that works with free text search, we consider it as the YouTube baseline.
On each experiment the participants compared our method with the two baselines and assigned a score for every combination of video from the first $k=5$ results and a selected topic name. The scores were binned to $s=0$: ``\textit{The topic is not included in the video}'', $s=1/2$: ``\textit{The topic is related with the video topics but not specific enough}'', and $s=1$: ``\textit{The topic is central to the video}''. 
The scores were average per participant, method, and task and considered as the $Perceived Precision@k$.

\begin{table}[ht]
\center
  \centering
    \begin{tabular}{p{0.3\linewidth} | p{0.3\linewidth} | p{0.23\linewidth}}
    \toprule
    Perceived P@k[$\uparrow$] & Task1 & Task2 \\
    \midrule
     TED/Search     & $0.47 \pm (0.30)$ & $0.21 \pm (0.22)$\\
    \hline
     YouTube         & $0.61 \pm (0.30) $& $0.18 \pm (0.20)$\\
    \hline
    Ours (Topics-Map)        & $\textbf{0.82}\pm (0.14)$ & $\textbf{0.67}\pm (0.21)$ \\
  \bottomrule
\end{tabular}
  \caption{Perceived Precision@k over \textsc{TED} comparing our method with YouTube and \textsc{TED} baselines.}
  \label{tab:UserStudy}
\end{table}

\textbf{Task \#1: Search over the \textsc{TED} ontology}\newline
The first task addressed the challenge in a multi-dimensional semantic search of multiple topics. The participants selected a set of 3-4 topics out of the TED ontology and asked to search for $k$ video names with these topics.
Users have evaluated whether the results align with the search topic names through its $Perceived Precision@k$.

We have found that our method significantly outperforms the baselines when it comes to $Precision@5=82\%$ ($P_{val} = 0.5\%$ in a single-factor Anova test). See comparison in Table~\ref{tab:UserStudy}.
When considered the top two methods, i.e., ours and YouTube, ours was found significantly better on a Pairwise t-Test ($P_{val} = 0.8\%$) even when applying the Bonferroni correction for multiple hypothesis testing.

\textbf{Task \#2: Fuzzy Search of open-set topics}\newline
The second task considered topics that are not included in the TED ontology. Participants were asked to search videos according to a pre-defined topics set.
We use the same baselines.

The Anova test have found a significant difference between methods in terms of $Precision@k$ with $P_{val}=3\cdot10^{-5}$.
A followup pairwise t-Test also revealed $Precision@k=67\%$ vs. $21\%$ ($P_{val} = 0.02\%$) for our method over the next best baseline (TED). Similarly, the Bonferroni correction holds.

\textbf{User Experience Evaluation - Interviews}\newline
The purpose of this study was to evaluate the effectiveness of our media archive browser, in supporting exploratory content discovery. To achieve this goal, we designed a questionnaire that assessed users' perceived value and satisfaction with our UX. 
The interviews have taught us that 8 out of 10 users voluntarily stated that the Topics-Map concept is intuitive and supports the query tuning process. Expert2 suggested using a Lasso selector to simultaneously include multiple locally related topic on the map.
Two participants took longer to gain confidence in the map and the concept of semantic similarity on a 2D plane.
A common feedback ($\frac{6}{10}$) was that the fact we always return results, while sometimes the baselines doesn't, makes our system much more productive as its perceived precision is high.

Expert1 was interested in Action Recognition Topics-Map and sub-video micro events for CCTV applications which identifies a domain gap from education content discovery.
Expert2 was asked how would the Topics-Map work for the university. He mentioned COVID as a major driver to online services like Zoom and says that although each course is self-contained, its true value should surface by expanding the search to the faculty level, university level, or even on a national scale of higher education. Quote:
``\textit{Students studying environmental science might not only be interested in lectures from their own course but also related lectures in biology, geology, or public policy. The content discovery system can help them easily navigate through and find relevant lectures from other courses that might have a connection to their primary area of study.}''

\section{Discussion and Limitations}

This paper presented a comprehensive solution for video archive browsing by leveraging AI models for media understanding, latent semantic embeddings, and a recommendation system to provide users with an interactive and efficient content discovery UX. 
We offered a Realtime video archive search experience that is scalable, robust, and easily adaptable to different domains and ontologies.

Our method relies on archive batch processing to generate semantic embeddings that are indexed to provide relevant search results efficiently. We highlighted that different domains require specific AI models to generate value from the data, and the choice of base-class networks ($86$M parameters) vs. foundation models (over $10$ trillion parameters) networks can have a significant impact on representation performance.
We also showed that text-based multimodality improves the representation quality for search purposes in both network classes over two datasets --- achieving SoTA results without finetuning.
Our Topics-Map concept was examined through a user study and interviews. The productivity of our method was found to significantly outperform both YouTube and the TED websites in two different AVS tasks: content discovery using TED ontology and custom topics ($P_{val} = 5\cdot10^{-3}\%$ and $P_{val} = 3\cdot10^{-5}\%$ respectively).
Overall, our proposed solution provides an effective and scalable video archive browsing method which is adaptable to various domains and applications.

The limitations of our approach are dependent on some of the architectural assumptions. Relying on archive batch indexing prior to any querying capability requires indexing also new videos which otherwise won't be found. 
Applying our method on different domains like CCTV or nature films, different set of AI models should be considered like action recognition and animal classifier.
Additionally, we assumed the availability of a topics ontology and labeled dataset in the case of supervised approach. Lastly, we discussed the practicality of using OpenAI as the representation method, as it may not always be possible due to network isolation or data privacy concerns. Therefore, applying a supervised approach with a pretrained model could be effective and efficient, as the model size impact operational costs and private data residency.

For future work, on the back-end side, we investigate generic architectures that may skip the textual fusion step by sharing tensor-based fusion from multiple AI models into a LLM decoder.
On the other hand, from the front-end perspective, we look into visualizations that could improve the abstraction of the Topics-Map. For example, by defining a metric which is inspired by TD-IDF to penalize on frequent topics like ``Science'', in the case of TED, which is too popular to be indicative or to run cluster analysis and unify related topics to reduce cognitive load.

\bibliographystyle{ACM-Reference-Format}
\bibliography{Main}










\end{document}